\documentclass[prd,twocolumn,superscriptaddress,preprintnumbers,showpacs]%
{revtex4}

\usepackage{epsfig}

\begin{document}

\title{Color Evaporation Description of
Inelastic Photo-Production of $J/\psi$ at HERA}

\author{O.\ J.\ P.\ \'Eboli}
\email{eboli@fma.if.usp.br}
\affiliation{Instituto de F\'{\i}sica, 
Universidade de S\~ao Paulo, S\~ao Paulo -- SP, Brazil.}

\author{E.\ M.\ Gregores}
\email{gregores@ift.unesp.br}
\affiliation{Instituto de F\'{\i}sica Te\'orica, 
Universidade Estadual Paulista, S\~ao Paulo -- SP, Brazil.}

\author{F.\ Halzen}
\email{halzen@pheno.physics.wisc.edu}
\affiliation{Department of Physics, 
University of Wisconsin, Madison -- WI, USA.}

\begin{abstract} 
  
The H1 Collaboration recently reported a new analysis on the inelastic
photo-production of $J/\psi$ mesons at DESY HERA $ep$ collider.  We
show that these new experimental results are well described by the
Color Evaporation Model for quarkonium production.  Moreover, this new
data requires the introduction of resolved photon contributions in
order to explain the results on small charmonium energy fraction,
indicating that colored $c \bar{c}$ pairs also contribute to the
process.

\end{abstract}

\preprint{\bf IFT-P.088/2002}
\preprint{\bf MADPH-02-1312}


\maketitle

\section{Introduction}

The H1 Collaboration recently reported an analysis on the inelastic
photo-production of $J/\psi$ mesons \cite{Adloff:2002ex} where their
new data was confronted with the color singlet \cite{Kramer:1995nb} and
color octet \cite{Kramer:2001hh} models.  We compare the H1 data with
the color evaporation model (CEM) predictions for quarkonium production
showing that it provides a good description of this new set of data.
Moreover, this new data probes for the first time the small charmonium
energy fraction ($z$) region, requiring the introduction of resolved
photon contributions in order to explain the results. These
contributions are basically due to color octet configurations,
confirming the correctness of their inclusion in the charmonium
production mechanism.

Previous measurements of the inelastic photo-production of charmonium
at HERA \cite{Aid:1996dn, Breitweg:1997we} appeared to have ignited a
charmonium crisis. The color singlet model (CSM) to onium production
cross section fits the data for large charmonium energy fraction $z$,
where color octet models seemed to fail.  This fact is, however, in
qualitative disagreement with a wealth of information that exists on
charmonium production by other initial states. We suggested
\cite{Eboli:1998xx} that this discrepancy is due to the neglect of
non-perturbative effects that are important at large $z$. Implementing
a phenomenological parametrization of these effects in a scheme
originally developed for Drell-Yan phenomenology, we illustrated how
agreement with data could be achieved. In this work, we employ the
procedure of \cite{Eboli:1998xx} to mimic the non-perturbative effects.

\section{Theoretical Framework} 

It is clear nowadays that $J/\psi$ production is a two--step process
where a heavy quark pair is produced first, followed by the
non--perturbative formation of the colorless asymptotic state. The
predictions of the once conventional treatment of color in perturbative
QCD calculations, {\em i.e.}, the CSM \cite{Baier:1983va}, for the
charmonium production at the Fermilab Tevatron is at variance with the
available data\cite{tev:psi_crisis}.  As a consequence, color octet as
well as singlet $c\bar{c}$ states contribute to the production of
$J/\psi$. Two formalisms have been proposed to incorporate these
features: the Non-Relativistic QCD, also known as COM
\cite{Bodwin:1994jh}, and the renewed CEM scheme \cite{Amundson:1995em,
Amundson:1996qr}, also known as Soft Color Interactions (SCI)
\cite{Edin:1997zb}.  The original CEM \cite{cem-old} actually predates
the color singlet approach, and had been abandoned for no good reason.
Recent measurements of the polarization of bound charm and beauty
mesons, seems to disfavor the COM framework \cite{Affolder:2000nn}.

The color evaporation model simply states that charmonium production
is described by the same dynamics as $D \bar{D}$ production, {\em
  i.e.}, by the formation of a colored $c\bar{c}$ pair.  Rather than
imposing that the $c\bar{c}$ pair is in a color-singlet state in the
short-distance perturbative diagrams, it is argued that the appearance
of color-singlet asymptotic states solely depends on the outcome of
large-distance fluctuations of quarks and gluons. These large-distance
fluctuations are probably complex enough for the occupation of
different color states to approximately respect statistical counting.
In other words, the formation of color-singlet states is a
non-perturbative phenomenon. In fact, it does not seem logical to
enforce the color-singlet property of the $c \bar{c}$ pair at short
distances, given that there is an infinite time for soft gluons to
readjust the color of the pair before it appears as an asymptotic
$\psi$, $\chi_c$ or, alternatively, $D \bar{D}$ state.  It is indeed
hard to imagine that a color-singlet state formed at a range
$m_{\psi}^{-1}$ automatically survives to form a $\psi$. This approach
to color is also used to formulate a successful prescription for the
production of rapidity gaps between jets at Tevatron
\cite{Eboli:1996gd, Eboli:1997ae, Eboli:1998ti, Eboli:1999dd,
  Eboli:fz} and HERA \cite{Eboli:1999dd, Eboli:fz, buch, Edin:1995gi}.

Although far more restrictive than other proposals, the CEM
successfully accommodates all features of charmonium production
\cite{Mariotto:2001sv, Kramer:2001hh, Schuler:1996ku}. It predicts
that the sum of the cross section of all onium and open charm states
is described by \cite{Amundson:1995em, Eboli:1996gd}
\begin{equation} 
   \sigma_{\rm onium} = \frac{1}{9}\int_{2 m_c}^{2 m_D} 
       dM_{c \bar{c}}~ \frac{d \sigma_{c \bar{c}}}{dM_{c\bar{c}}} \; , 
\label{sig:on} 
\end{equation} 
and 
\begin{equation} 
  \sigma_{\rm open} = \frac{8}{9}  \int_{2 m_c}^{2 m_D} 
    dM_{c\bar{c}}~\frac{d \sigma_{c\bar{c}}}{d M_{c \bar{c}}} 
   + \int_{2 m_D} dM_{c\bar{c}}~\frac{d\sigma_{c\bar{c}}}{dM_{c\bar{c}}} \; , 
\label{sig:op} 
\end{equation} 
where $M_{c\bar{c}}$ is the invariant mass of the $c\bar c$ pair. The
factor $1/9$ stands for the probability that a pair of charm quarks
formed at a typical time scale $1/M_\psi$ ends up as a color singlet
state after exchanging an uncountable number of soft gluons with the
reaction remnants.  One attractive feature of this model is the
relation between the production of charmonium and open charm which
allows us to use the open charm data to normalize the perturbative QCD
calculation, and consequently, to make more accurate predictions for
charmonium cross sections.

The fraction $\rho_\psi$ of produced onium states that materialize as
$\psi$,
\begin{equation}
       \sigma_\psi = \rho_\psi~\sigma_{\rm onium} \; ,
\label{frac}
\end{equation}
has been inferred from low energy measurements to be a constant
\cite{gavai,schuler}. From the charmonium photo-production, we
determined that $\rho_\psi=0.43$--0.5 \cite{Amundson:1996qr}; a value
that can be accounted for by statistical counting of final states
\cite{Edin:1997zb}. The fact that all $\psi$ production data are
described in terms of this single parameter, fixed by $J/\psi$
photo-production, leads to parameter free predictions for $Z$-boson
decay rate into prompt $\psi$ \cite{Gregores:1996ek}, and to
charmonium production cross section at Tevatron
\cite{Eboli:1999hh} and HERA \cite{Eboli:1998xx}, as well as in
neutrino initiated reactions \cite{Eboli:2001hc}. These predictions
are in agreement with the available data.

\section{Inelastic Photo-production of Charmonium}

Application of the CEM scheme to inelastic charmonium
photo-production is straightforward. 
According to the parton model, the cross section for the $J/\psi$
photo-production at a given center-of-mass energy $W$ is
\begin{widetext}
\begin{equation}
\sigma_{\gamma p\rightarrow J/\psi X}(W)=
\int\!\!\int f_{A/\gamma}(x_A) \, f_{B/p}(x_B) 
\,\hat\sigma_{AB\rightarrow J/\psi X}(\hat s) \, dx_A \, dx_B
\; ,
\end{equation}
\end{widetext}
where the subprocess cross section $\hat\sigma$ is given by Eqs.\ 
(\ref{sig:on}) and (\ref{frac}).  Here, $\sqrt{\hat s}=\sqrt{x_A x_B}
\, W$ is the center-of-mass energy of the subprocess $AB\rightarrow
J/\psi X$, and $f_{A/\gamma}$ ($f_{B/p}$) is the distribution function
of the parton $A$ ($B$) in the photon (proton).  For direct photon
interactions $(A=\gamma)$ we have $f_{A/\gamma}(x_A)=\delta(x_A-1)$.

An important kinematical variable is 
\begin{equation}
z \equiv \frac{P_{J/\psi}\cdot P_{p}}{P_{\gamma}\cdot P_{p}} \; ,
\end{equation}
where $P_{J/\psi,\gamma,p}$ is the four-momentum of the $J/\psi$,
photon, or proton, respectively. In the proton rest frame, $z$ is the
fraction of photon energy carried by the $J/\psi$.

\begin{figure}[t!]
\epsfig{file=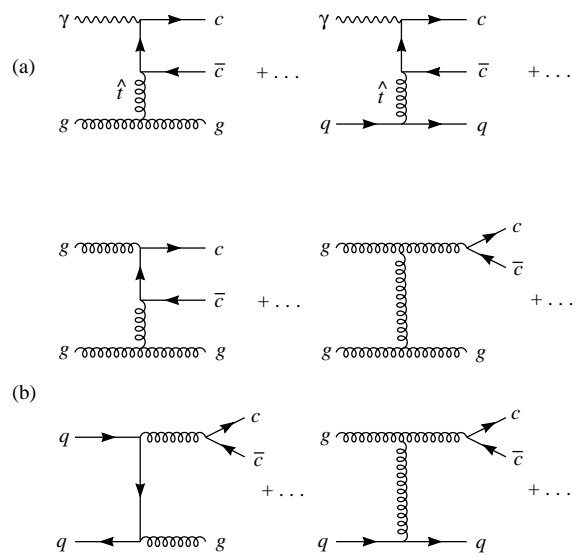,width=0.9\linewidth,clip=}
\caption{\label{fig:feyn}
  Processes contributing to direct (a) and resolved (b) charmonium
  photo-production.}
\end{figure}

The lowest order CEM contribution for the charmonium production is the
direct photon process $\gamma g \rightarrow c \bar{c}$. However, it is
important only for $z\approx 1$.  For the range of $z$ we are
interested, the direct photon contribution is dominated by the
diagrams depicted in the Fig.~\ref{fig:feyn}(a). The charm quark pair
in $\gamma g$ fusion can be produced in both color singlet and octet
configurations, while $\gamma q$ fusion leads only to colored $c
\bar{c}$ pairs.  Besides the direct photon--gluon and photon--quark
reactions, we also included resolved photon processes, which proceed
via quark--quark, quark--gluon, and gluon--gluon fusion into $c
\bar{c}$+quark (gluon), as exemplified in Fig.~\ref{fig:feyn}(b).

The higher order processes, like the ones in Fig.~\ref{fig:feyn}, have
to be evaluated with some caution in the region of relatively small
$J/\psi$ transverse momentum and large $z$. For small $t$-channel
momentum transfer ($\hat{t}\rightarrow 0$), the gluon exchange
diagrams in Fig.~\ref{fig:feyn} represent the QCD evolution of the
initial state gluon distribution functions, and not higher order gluon
exchange diagrams.  Therefore, the contribution of the gluon-exchange
diagrams where $\hat t$ is soft, has already been taken into account
by the leading order $\gamma~ g \rightarrow c~ \bar{c}$ diagram
\cite{nas88}. By itself, when evaluated at tree level, the diagrams in
Fig.~\ref{fig:feyn} lead to a divergence for $z\rightarrow 1$, that
appears as an unphysical growth of the cross section for $z \lesssim
1$.  Although the double counting is relevant only for vanishing $p_T$
and $z=1$, the enhancement of the cross section associated with the
collinear divergence will, in practice, affect the calculation until
$p_T$ and $z$-values become truly perturbative.

Presently, complete higher order QCD calculations are not available
for the processes we are interested in. Therefore, we
phenomenologically took into account non-perturbative QCD corrections
in large $z$ charmonium photo-production by introducing the following
parametrization to mimic the non-perturbative contributions
\cite{Eboli:1998xx}:
\begin{equation}
\frac{d^2\sigma}{dp_Tdz}=
[1-F(Q_0,p_T)][1-G(Q_0,z)]\frac{d^2\sigma_{tree}}{dp_Tdz}
\label{phen}
\end{equation}
with
\begin{equation}
F(Q_0,p_T)=e^{-\frac{p_T^2}{k_T^2}}
\end{equation}
and
\begin{equation}
G(Q_0,z)=e^{-\frac{1-z}{z_0z}} \, ,
\end{equation}
where $\sigma_{tree}$ is the tree level perturbative cross section,
$k_T^2 = z (Q_0^2 + 4m_c^2) - 4m_c^2$, and $(1-z_0) =
(p_T^2+4m_c^2)/(Q_0^2+4m_c^2)$ are positive definite or null.  The
parameter $Q_0$ indicates the value of the momentum transfer $\hat t$
where the higher order diagrams contribute in the truly perturbative
regime \cite{berg81}. We found in \cite{Eboli:1998xx} that $Q_0=2m_c$
best describes the data, and it is this value that we used in the
present work.

\section{Results}

In this work we evaluated numerically the tree level scattering
amplitudes using the Madgraph \cite{madgraph} and Helas \cite{helas}
packages.  The phase space integration was performed using the
adaptative Monte Carlo program VEGAS \cite{vegas}.

There is a wide choice of QCD parameters appearing in the evaluation
of charmonium photo-production, which leads to a large theoretical
uncertainty. For the sake of definiteness, we present only the
uncertainty that arises from possible choices of the charm quark mass,
keeping all others QCD parameters fixed. All our results are presented
as a central value surrounded by an error band corresponding to the
choice of the charm quark mass as $m_c = 1.3 \pm 0.1$ GeV. We used the
GRV-94 LO \cite{Gluck:1994uf} parameterization of the proton structure
functions and GRV-G LO \cite{Gluck:1994tv} for the photon parton
density. For both structure functions, we set the factorization scale
as $\mu_F = \sqrt{\hat s}$. We evaluated the running of the strong
coupling constant in leading order with four active flavors using
$\Lambda_{QCD}=300$ MeV, renormalization scale $\mu_R=\sqrt{2}m_c$ for
direct processes, and $\mu_R=\sqrt{\hat s}$ for resolved processes. We
also used $\rho_\psi=0.5$ as the $J/\psi$ fraction of all charm bound
states.

The H1 Collaboration performed their analysis using data in which
$Q^2<1$ GeV$^2$, and subdivided their data into several different
kinematical regions, in order to better determine the region where
perturbative QCD calculations furnish a reliable description of the
data.

\begin{figure}[h]
\epsfig{file=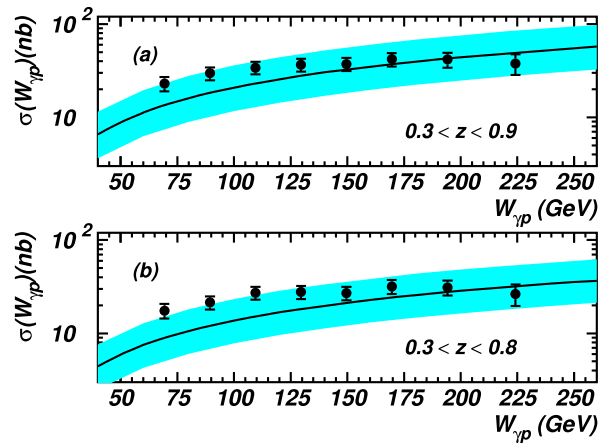,width=0.9\linewidth,clip=}
\caption{\label{fig:xsec}
  Total cross section as function of $W_{\gamma p}$ for $p_T>1$ GeV:
  (a) $0.3<z<0.9$ and (b) $0.3<z<0.8$. The shaded band shows the
  theoretical prediction for $m_c = 1.3 \pm 0.1$ GeV.}
\end{figure}
We collated the CEM predictions with the H1 results using the cuts
applied to the experimental data.  Initially we analyzed the behavior
of the total cross section as function of the $\gamma p$
center--of--mass energy ($W$) for two different $z$ regions, requiring
a minimum $J/\psi$ transverse momentum $p_T>1$ GeV. In
Fig.~\ref{fig:xsec}(a) we show the CEM predictions for $0.3<z<0.9$,
and in Fig.~\ref{fig:xsec}(b) we present our results for $0.3<z<0.8$.
As we can see from these figures, moving the upper limit in $z$ alone
does not change much the shape of the curve, and within theoretical
uncertainties, agreement is found for both regions.

\begin{figure}
\epsfig{file=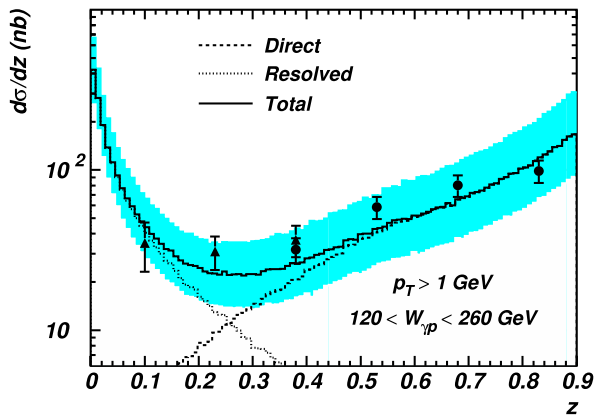,width=0.9\linewidth,clip=}
\caption{\label{fig:zall}
  Differential cross section as function of the inelasticity parameter
  $z$ for $p_T^2 > 1$ GeV$^2$ and $120<W_{\gamma p}<260$ GeV. The
  dashed line stands for the contribution from direct processes while
  the dotted line shows the contribution from resolved processes. The
  solid line is the sum of both contributions. The shaded band shows
  the theoretical uncertainty on the value of the charm mass ($m_c =
  1.3 \pm 0.1$ GeV).}
\end{figure}

We display in Fig.~\ref{fig:zall}, the CEM predictions for the $z$
spectrum, requiring $p_T>1$ GeV and a reaction center--of--mass energy
in the range $120<W_{\gamma p}<260$ GeV. In this figure, the
experimental points represented by triangles and squares stand for two
different datasets; see Ref.\ \cite{Adloff:2002ex} for details. The
direct photon contribution is represented by the dashed line, and the
resolved one by the dotted line; the solid line displays the sum of
direct and resolved contributions. The shaded area corresponds to the
total CEM prediction for the charm quark mass described above. As we
can see from this figure, the agreement of the CEM with data is quite
good. Moreover, the data at low and medium $z$ do require the
introduction of resolved processes in order to explain the results.
This is a clear signal that colored charm quark pairs contribute to
the $J/\psi$ production.

\begin{figure}[h]
\epsfig{file=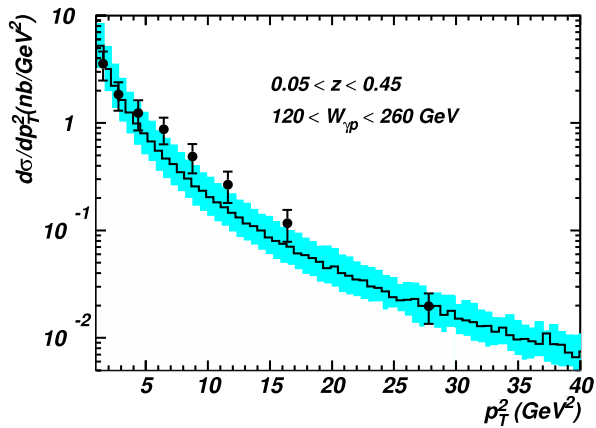,width=0.9\linewidth,clip=}
\caption{\label{fig:ptlow}
  Squared transverse momentum $p_T^2$ distribution for data collected
  in the very inelastic region $0.05<z<0.45$ and $120<W_{\gamma
    p}<260$ GeV.  The shaded band shows the theoretical uncertainty on
  the value of the charm mass ($m_c = 1.3 \pm 0.1$ GeV).}
\end{figure}

In Fig.~\ref{fig:ptlow} we present the $p_T^2$ distribution for the low
$z$ sample $0.05<z<0.45$ which corresponds to the triangles in
Fig.~\ref{fig:zall}. For these values of $z$ we expect the theoretical
uncertainties due to higher order corrections to be small, and we can
see that the CEM and data agree well. Consequently, the
disagreement between data and theory at large $z$ is indeed due to the
importance of higher order non--perturbative corrections.

Higher order effects are sizeable on the medium and high $z$
regions.  Therefore, the H1 Collaboration divided their data sample into
several $p_T$ and $z$ regions for a center--of--mass energies in the
range $60<W_{\gamma p}<240$ GeV. This allows us to better compare
data with theory.

\begin{figure}[h]
\epsfig{file=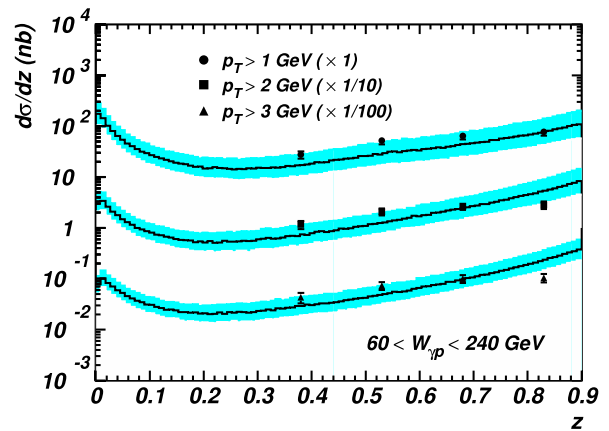,width=0.9\linewidth,clip=}
\caption{\label{fig:zmid}
  Differential cross section as function of the inelasticity parameter
  $z$ for $p_T > 1,\, 2,$ and $3$ GeV, divided by a factor $1,\, 10,$
  and $100$ respectively. The shaded bands show the theoretical
  prediction obtaining varying the charm mass ($m_c = 1.3 \pm 0.1$
  GeV).}
\end{figure}

In Fig.~\ref{fig:zmid}, we compare the CEM predictions for the $z$
spectrum with data for three different values of the $J/\psi$ minimum
transverse momentum, \textit{i.e.}, $p_T>1,\, 2,$ and $3$ GeV. The
curves have been divided by factors $1,\, 10,$ and $100$, respectively,
in order to help visualization. There is an overall agreement between
data and theory, except for the highest bin in $z$. Moreover, removing
our regularization procedure only worsens the theoretical results
\cite{Eboli:1998xx}.  Increasing the minimum value of the transverse
momentum does not improve the quality of the fitting, what means that
our parametrization of the higher order effects has correctly
incorporated the dependence on minimum $p_T$.

\begin{figure}[h]
\epsfig{file=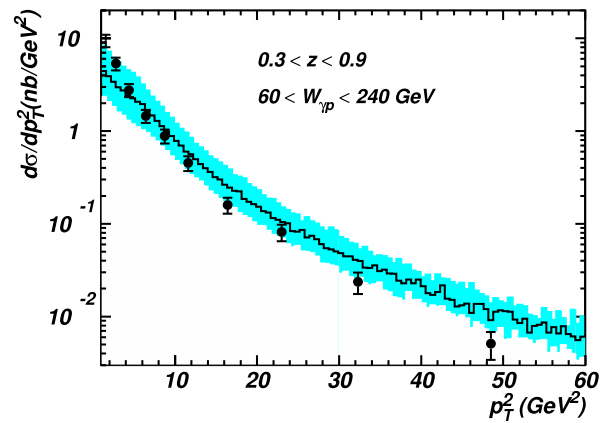,width=0.9\linewidth,clip=}
\caption{\label{fig:ptmid}
  Squared transverse momentum $p_T^2$ distribution for $0.3<z<0.9$ and
  center-of-mass energy $60<W_{\gamma p}<240$ GeV.  The shaded band
  shows the theoretical uncertainty on the value of the charm mass
  ($m_c = 1.3 \pm 0.1$ GeV).}
\end{figure}

\begin{figure}[t]
\epsfig{file=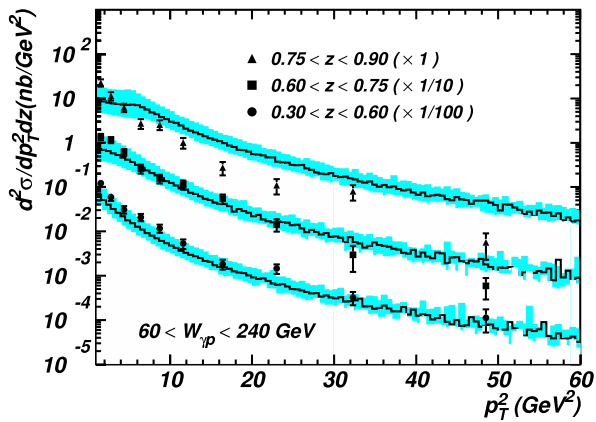,width=0.9\linewidth,clip=}
\caption{\label{fig:dptdz}
  Squared transverse momentum $p_T^2$ spectrum for $60<W_{\gamma
    p}<240$ and data collected in three different inelastic bins:
  $0.75<z<0.9$ (upper), $0.6<z<0.75$ (middle), and $0.3<z<0.6$
  (lower). To help visualization the curves have been scaled as 1,
  1/10, and 1/100 from top to bottom.  The shaded bands show the
  theoretical uncertainty on the value of the charm mass ($m_c = 1.3
  \pm 0.1$ GeV).}
\end{figure}

Fig.~\ref{fig:ptmid} contains the $J/\psi$ squared transverse momentum
distribution taking into account the medium and high $z$ data. The
agreement between theory and data is quite satisfactory. However, the
shape of the spectrum shows some disagreement at the very low $p_T$
bins. In order to understand what is happening, let us consider this
distribution for three different $z$ bins, namely $0.75<z<0.9$,
$0.6<z<0.75$, and $0.3<z<0.6$; see Fig.~\ref{fig:dptdz} where the
curves have also been divided by factors $1,\,10,$ and $100$,
respectively, to help visualization. We can learn from this last
figure that the agreement between CEM predictions and data improves
for low $z$ regions.  This fact can be understood as a limitation of
the proposed parameterization to completely mimic higher order QCD
contributions when we approach the elastic region. It also implies
that the observed small discrepancy with data for the highest bin on
$z$ must be credited to the lack of a complete QCD calculation and is
not related to the Color Evaporation approach to describe quarkonium
production.

\section{Conclusions}

We showed that the Color Evaporation Model describes the available data
on $J/\psi$ photo--production, provide we include high order QCD
corrections at high inelasticities $z$. Moreover, the newly available
data at low $z$ provides a clear proof of the importance of colored $c
\bar{c}$ pairs to the production of charmonium, since the data on this
region can only be explained considering resolved photon processes,
which lead to colored $c \bar{c}$ pairs.

\bigskip

\begin{acknowledgments}
  This research was supported in part by the University of Wisconsin
  Research Committee with funds granted by the Wisconsin Alumni
  Research Foundation, by the U.S.\ Department of Energy under grant
  DE-FG02-95ER40896, by Funda\c{c}\~{a}o de Amparo \`a Pesquisa do
  Estado de S\~ao Paulo (FAPESP), by Conselho Nacional de
  Desenvolvimento Cient\'{\i}fico e Tecnol\'ogico (CNPq), and by
  Programa de Apoio a N\'ucleos de Excel\^encia (PRONEX).
\end{acknowledgments}



\begin{thebibliography}
\frenchspacing

\bibitem{Adloff:2002ex}
C.~Adloff {\it et al.}  [H1 Collaboration],
DESY-02-059, May 2002 [hep-ex/0205064].

\bibitem{Kramer:1995nb}
M.~Kramer,
Nucl.\ Phys.\ B {\bf 459}, 3 (1996).

\bibitem{Kramer:2001hh}
M.~Kramer,
Prog.\ Part.\ Nucl.\ Phys.\  {\bf 47}, 141 (2001).

\bibitem{Aid:1996dn}
S.~Aid {\it et al.}  [H1 Collaboration],
Nucl.\ Phys.\ B {\bf 472}, 3 (1996).

\bibitem{Breitweg:1997we}
J.~Breitweg {\it et al.}  [ZEUS Collaboration],
Z.\ Phys.\ C {\bf 76}, 599 (1997).

\bibitem{Eboli:1998xx}
O.~J.~Eboli, E.~M.~Gregores, and F.~Halzen,
Phys.\ Lett.\ B {\bf 451}, 241 (1999).

\bibitem{Baier:1983va}
R.~Baier and R.~Ruckl,
Z.\ Phys.\ C {\bf 19}, 251 (1983).

\bibitem{tev:psi_crisis}
F.~Abe {\it et al.} [CDF Collaboration],
Phys. Rev. Lett. {\bf 69}, 3704 (1992);
{\it ibid} {\bf 79}, 572 (1997);
{\it ibid} {\bf 79}, 578 (1997);
S.~Abachi {\it et al.} [D\O\ Collaboration],
Phys. Lett. B {\bf 370}, 239 (1996).

\bibitem{Bodwin:1994jh}
G.~T.~Bodwin, E.~Braaten, and G.~P.~Lepage,
Phys.\ Rev.\ D {\bf 51}, 1125 (1995)
[Erratum-ibid.\ D {\bf 55}, 5853 (1997)].

\bibitem{Amundson:1995em}
J.~F.~Amundson, O.~J.~Eboli, E.~M.~Gregores, and F.~Halzen,
Phys.\ Lett.\ B {\bf 372}, 127 (1996).

\bibitem{Amundson:1996qr}
J.~F.~Amundson, O.~J.~Eboli, E.~M.~Gregores, and F.~Halzen,
Phys.\ Lett.\ B {\bf 390}, 323 (1997).

\bibitem{Edin:1997zb}
A.~Edin, G.~Ingelman, and J.~Rathsman,
Phys.\ Rev.\ D {\bf 56}, 7317 (1997).

\bibitem{cem-old}
H.~Fritzsch, Phys.\ Lett.\ B {\bf 67}, 217 (1977); 
F.~Halzen, Phys.\ Lett.\ B {\bf 69}, 105 (1977); 
F.~Halzen and S.~Matsuda, Phys.\ Rev.\ D {\bf 17}, 1344 (1978).

\bibitem{Affolder:2000nn}
T.~Affolder {\it et al.}  [CDF Collaboration],
Phys.\ Rev.\ Lett.\  {\bf 85}, 2886 (2000).

\bibitem{Mariotto:2001sv}
C.~B.~Mariotto, M.~B.~Gay Ducati, and G.~Ingelman,
Eur.\ Phys.\ J.\ C {\bf 23}, 527 (2002).

\bibitem{Schuler:1996ku}
G.~A.~Schuler and R.~Vogt,
Phys.\ Lett.\ B {\bf 387}, 181 (1996).

\bibitem{Eboli:1996gd}
O.~J.~Eboli, E.~M.~Gregores, and F.~Halzen,
{\it Proceedings of the 26th International Symposium on 
Multiparticle Dynamics (ISMD\,96)}, Faro, Portugal, 1996.

\bibitem{Eboli:1997ae}
O.~J.~Eboli, E.~M.~Gregores, and F.~Halzen,
Nucl.\ Phys.\ Proc.\ Suppl.\  {\bf 71}, 349 (1999).

\bibitem{Eboli:1998ti}
O.~J.~Eboli, E.~M.~Gregores, and F.~Halzen,
Phys.\ Rev.\ D {\bf 58}, 114005 (1998).

\bibitem{Eboli:1999dd}
O.~J.~Eboli, E.~M.~Gregores, and F.~Halzen,
Phys.\ Rev.\ D {\bf 61}, 034003 (2000).

\bibitem{Eboli:fz}
O.~J.~Eboli, E.~M.~Gregores, and F.~Halzen,
Nucl.\ Phys.\ Proc.\ Suppl.\  {\bf 99A} 257 (2001) .

\bibitem{buch}
W.~Buchm\"uller,
Phys.\ Lett.\ B {\bf 353}, 335 (1995); 
W.~Buchm\"uller and A.~Hebecker,
Phys.\ Lett.\ B {\bf 355}, 573 (1995).

\bibitem{Edin:1995gi}
A.~Edin, G.~Ingelman, and J.~Rathsman,
Phys.\ Lett.\ B {\bf 366}, 371 (1996).

\bibitem{gavai}
R.~Gavai {\em et al.},
Int.\ J.\ Mod.\ Phys.\ A {\bf 10}, 3043 (1995).

\bibitem{schuler}
G.~Schuler, report CERN-TH.7170/94.

\bibitem{Gregores:1996ek}
O.~J.~Eboli, E.~M.~Gregores, and F.~Halzen, 
Phys.\ Lett.\ B {\bf 395}, 113 (1997).

\bibitem{Eboli:1999hh}
O.~J.~Eboli, E.~M.~Gregores, and F.~Halzen,
Phys.\ Rev.\ D {\bf 60}, 117501 (1999).

\bibitem{Eboli:2001hc}
O.~J.~Eboli, E.~M.~Gregores, and F.~Halzen,
Phys.\ Rev.\ D {\bf 64}, 093015 (2001).

\bibitem{nas88}
P.~Nason, S.~Dawson, and R.~K.~Ellis, 
Nucl. Phys. B {\bf 303}, 607 (1988).

\bibitem{berg81}
E.~Berger and D.~Jones, Phys. Rev. D {\bf 23}, 1521 (1981).

\bibitem{madgraph}
W.~Long and T.~Steltzer,
Comput.\ Phys.\ Commun.\ {\bf 81}, 357 (1994).

\bibitem{helas}
H.~Murayama, I.~Watanabe, and K.~Hagiwara, KEK report 91-11.

\bibitem{vegas}
G.~P.~Lepage,
CLNS-80/447.

\bibitem{Gluck:1994uf}
M.~Gluck, E.~Reya, and A.~Vogt,
Z.\ Phys.\ C {\bf 67}, 433 (1995).


\bibitem{Gluck:1994tv}
M.~Gluck, E.~Reya, and M.~Stratmann,
Phys.\ Rev.\ D {\bf 51}, 3220 (1995).

\end{thebibliography}
\end{document}